\documentclass[a4paper,11pt]{article}
\usepackage{pos}
\usepackage{wrapfig}

\title{The strong coupling from the IR to the UV extremes: Determination of $\alpha_s$ and prospects from EIC and JLab at 22 GeV}
 \ShortTitle{$\alpha_s$ at EIC and JLab@22}

\author*[a]{Alexandre Deur}

\affiliation[a]{Thomas Jefferson National Accelerator Facility,\\
12000 Jefferson Avenue  NewportNews, VA 23606, USAn}

\emailAdd{deurpam@jlab.org}

\abstract{We discuss how the Bjorken sum rule allows access to the QCD running coupling $\alpha_s$ at any scale, including in the deep infrared IR domain. The Bjorken sum data from Jefferson Lab, together with the world data on $\alpha_s$ reported by the Particle Data Group, allow us to determine the running of $\alpha_s(Q)$ over five orders of magnitude in four-momentum $Q$. We present two possible future measurements of the running of $\alpha_s(Q)$ using the Bjorken sum rule: the first at the EIC, covering the range $1.5 < Q < 8.7$ GeV, and the second at Jefferson Lab at 22 GeV, covering the range $1.0 < Q < 4.7$ GeV.}

\FullConference{
}

\begin{document}
\maketitle

\section{Introduction}
\vspace{-3mm}
The strong coupling $\alpha_s$ sets the magnitude of the strong interaction. Consequently, it is the central quantity of QCD 
and an essential parameter of the Standard Model~\cite{Gross:2022hyw, Deur:2023dzc}. 
However, the current experimental accuracy on $\alpha_s$, $\Delta \alpha_s/\alpha_s = 0.85\%$~\cite{ParticleDataGroup:2024cfk}, 
makes it the least well known of the fundamental couplings.  
To compare, $\Delta \alpha/\alpha=1.5\times 10^{-10}$ for QED, 
$\Delta G_F/G_F=5.1\times 10^{-7}$ for the weak force,
and $\Delta G_N/G_N=2.2\times 10^{-5}$ for gravity. 
This relative lack of precision limits the studies of the strong force in the perturbative QCD (pQCD) domain,
hinders pQCD tests and searches for physics beyond the Standard Model, and, at low energy, 
impedes the study of nonperturbative approaches to QCD. Therefore, large efforts are ongoing to reduce $\Delta \alpha_s/\alpha_s$~\cite{dEnterria:2022hzv}.

No known single experiment can exquisitely determine $\alpha_s$. 
Currently, the best individual experimental determinations reach only the $\sim1\%-2\%$ level. 
Thus, the strategy is to combine many independent results to achieve the desired precision of $\sim$$0.1\%$~\cite{dEnterria:2022hzv}. 
One method to access $\alpha_s$ is by using deep inelastic scattering (DIS) {\it via} the momentum-evolution of 
observables. For example, one may fit  $g_1(x,Q^2)$, the nucleon longitudinal spin structure function. 
(Here, $Q^2$ is the 4-momentum transfer in the inclusive lepton scattering used to measure $g_1$, and $x$ is the Bjorken scaling variable.)
Fitting $g_1(x,Q^2)$ is a complex endeavor that involves DGLAP~\cite{Gribov:1972ri} 
global fits and modeling nonperturbative inputs, namely quark/gluon Parton Distribution Functions (PDF) and Higher-Twists (HT) if the data cover 
low-$Q^2$/high-$x$. 
An alternative is to fit the $Q^2$ evolution of the $g_1$ moment, $\Gamma_1 \equiv \int_0^1 g_1 dx$. With the $x$-dependence 
integrated out, the formalism simplifies. 
Furthermore, modeling PDFs is not needed since the nonperturbative inputs are 
the measured axial charges $a_0$, $a_8$ and 
$g_A$~\cite{Deur:2018roz}. However, other issues arise. One is the unreachable low-$x$ part of the moment.
(How low in $x$ an experiment can reach depends on the beam energy, how forward the measurement is carried out and the minimum $Q^2$ value tolerable.)
Also, $a_0$ depends on $Q^2$ and may receive a contribution from the poorly known polarized gluon PDF, depending on the chosen  renormalization scheme (RS). 
A major  simplification occurs by considering the isovector part of the moment, 
$\int g_1^p-g_1^ndx \equiv \Gamma_1^{p-n}$, i.e., the integral involved in the Bjorken sum rule (BJSR)~\cite{Bjorken:1966jh}. 
($p$ and $n$ denote the proton and neutron.) 
$\Gamma_1^{p-n}$ has a simple $Q^2$ evolution, known to a order higher than in the single-nucleon case, 
which is crucial since pQCD truncation errors are typically the main limitation on precise extractions of $\alpha_s$~\cite{dEnterria:2022hzv}. 
In addition, the main nonperturbative input is the precisely measured $g_A$
~\cite{ParticleDataGroup:2024cfk}
--no gluon PDF is needed-- 
and HT are known to be small for $\Gamma_1^{p-n}$~\cite{Deur:2014vea}.

The BJSR $\overline{\rm MS}$ approximant at N$^4$LO~\cite{Kataev:1994gd}, with a N$^5$LO estimate~\cite{Bjorken_a5} and a twist-4 term is:
\setlength{\abovedisplayskip}{1pt}
\setlength{\belowdisplayskip}{2pt}
\begin{eqnarray}
\Gamma_1^{p-n} = \frac{1}{6}g_A  \bigg[1- \frac{\alpha_s}{\pi}-3.58\bigg(\frac{\alpha_s}{\pi}\bigg)^2-20.215\bigg(\frac{\alpha_s}{\pi}\bigg)^3 
 - 175.7\bigg(\frac{\alpha_s}{\pi}\bigg)^4- \sim893\bigg(\frac{\alpha_s}{\pi}\bigg)^5 \bigg] 
+ \frac{\mu_4}{Q^2}.
\label{eq:BJSR}
\end{eqnarray}
(The coefficients are $n_f$ dependent. We provided them for $n_f=3$.)
One can then extract $\alpha_s$ from $\Gamma_1^{p-n}$ in two ways. First, for each data point, one may solve Eq.~(\ref{eq:BJSR}) for $\alpha_s$. This maps the
$Q^2$ evolution of $\alpha_s$ but $\Delta \alpha_s/ \alpha_s$ increases quickly with $Q^2$.
Second, one may extract $\alpha_s$ from the overall $Q^2$-dependence of $ \Gamma_1^{p-n}$. This relative method is more accurate
but provides only one value of $\alpha_s$. Both methods will be discussed here. 
We will show first how the point-by-point method has produced a $Q^2$ mapping of $\alpha_s$ at low $Q^2$, complementing the dataset in the pQCD domain. 
Together, the two datasets provide $\alpha_s$ at essentially all scales. 
Then, we will show that measuring the BJSR at the Electron Ion Collider 
(EIC)~\cite{Accardi:2012qut} and with Jefferson Lab (JLab) upgraded to 22 GeV (JLab@22)~\cite{Accardi:2023chb} 
offers good prospects for mapping 
$\alpha_s(Q)$ and for precisely determining $\alpha_s(M_z)$ at
$Q=M_z$, the $Z^0$ mass at which the value of $\alpha_s$ is usually quoted. 

\section{The running of $\alpha_s$ at all scales }
\begin{figure}
\center
\includegraphics[width=0.55\textwidth]{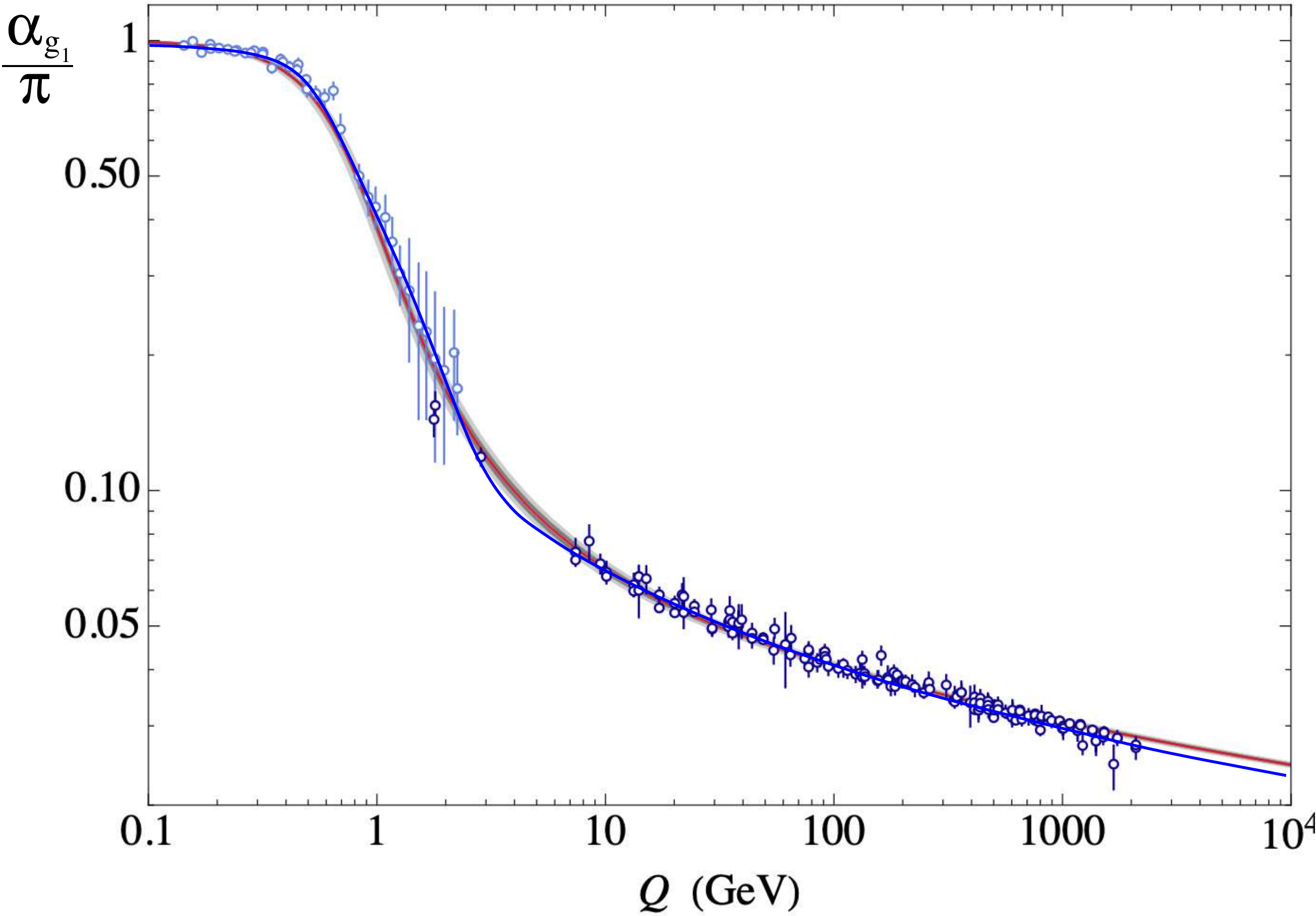}
\caption{\small{$\alpha_{g_1}/\pi$ at all scales. 
The low $Q$ data (light blue) are from CERN, DESY, JLab and SLAC~\cite{Deur:2021klh},
while the large $Q$ dataset (dark blue) is the PDG compilation~\cite{ParticleDataGroup:2024cfk} transformed to the $g_1$ scheme.
The red line is the prediction~\cite{deTeramond:2024ikl} incorporating AdS/QCD constraints at low $Q$ and pQCD ones at large $Q$.
The blue line is the simple fit form $aT_r/\ln\big([Q^2+Q^2_r]/\Lambda^2\big)$ with $a$=1.56, $T_r$=$\big(1+(\pi-1)/(e^{(Q-f)/g}+1)\big)$,
$Q_r$=$b/\big(e^{(Q^2-c)/d} +1\big)$, $\Lambda$=0.246~GeV, $b=$0.808~GeV, $c$=0.11~GeV$^2$, $d$=0.20~GeV$^2$, $f$=1.29~GeV 
and $g$=0.59~GeV~\cite{Deur:2025rjo}.}}
 \label{fig:alpha_s-all-scales}
\end{figure}

\vspace{-3mm}
While the definition and determination of $\alpha_s$ in the pQCD domain is well established~\cite{dEnterria:2022hzv, ParticleDataGroup:2024cfk},
it has not been so in the nonperturbative regime. However, some definitions of $\alpha_s$ allow it to be determined also in this regime,
enabling investigations of $\alpha_s$ there with both experiments and theory. In fact, since $\alpha_s$ is not an observable, 
different definitions are possible~\cite{Deur:2023dzc, Deur:2016tte}. In particular, the ``effective charge'' prescription~\cite{Grunberg:1980ja} 
allows for a definition applicable at any scale. It defines $\alpha_s$ from an observable's pQCD approximant truncated to its LO in $\alpha_s$. 
In that context, a well-suited observable is the Bjorken sum. Using the effective charge prescription, Eq.~(\ref{eq:BJSR}) becomes 
$ \Gamma_1^{p-n}(Q^2)=  \frac{g_A}{6}~\big[1-\frac{\alpha_{g_1}(Q^2)}{\pi}\big]$, where the notation $\alpha_{g_1}$ 
signals the chosen observable, a choice equivalent to adopting a particular RS~\cite{Deur:2016cxb}. 
The effective charge prescription folds into $\alpha_s$ both short-distance pQCD effects from DGLAP and long-distance effects
(e.g., HT), generalizing the procedure that transmutes 
a coupling {\it constant} into a {\it running} effective coupling. 
Effective charges have advantages: they are extractable at any 
scale, free of Landau p\^ole, have improved pQCD series convergence and are RS-independent. The latter 
comes from the RS-independence of the LO coefficient of any pQCD series. On the other hand, an
effective charge depends on the defining process. Yet, QCD predictability is preserved  
since different types of effective charges can be related~\cite{Brodsky:1994eh}. 
The Bjorken sum is particularly suited to define an effective charge thanks to the advantages already mentioned 
(simple pQCD series assessed up to $\alpha_s^5$, small-to-vanishing coherent process contributions such as resonance scattering or HT).
Furthermore, $\Gamma_1^{p-n}$ data exist at low, intermediate, and high $Q^2$ and can be supplemented by 
rigorous sum rules dictating its behavior in the 
unmeasured $Q^2 \to 0$ (Gerasimov-Drell-Hearn (GDH) sum rule~\cite{Gerasimov:1965et}) and 
$Q^2 \to \infty$ (BJSR) limits. Consequently, $\alpha_{g_1}$ is accurately known at any $Q^2$
(Fig.~\ref{fig:alpha_s-all-scales}). 
The $Q^2 \lesssim 10$~GeV$^2$ data come from CERN, DESY, JLab and SLAC, see Refs.~\cite{Deur:2021klh} for the latest data on 
$\Gamma_1^{p-n}$ and~\cite{Deur:2022msf} for the latest extraction of $\alpha_{g_1}$. The higher $Q^2$ data in Fig.~\ref{fig:alpha_s-all-scales}
are  the PDG compilation~\cite{ParticleDataGroup:2024cfk} transformed from the ${\rm \overline{MS}}$ RS to the $g_1$ scheme. 
At large $Q^2$, data and predictions agree, reflecting the consistency of the effective charge prescription and pQCD. At low $Q$, 
the data agree well with the predictions using the effective charge definition, namely 
AdS/QCD~\cite{Brodsky:2010ur, deTeramond:2024ikl} and Schwinger-Dyson Equation/lattice QCD~\cite{Binosi:2016nme} calculations. 
The agreement 
is all the more remarkable since the parameters in calculations~\cite{Brodsky:2010ur, deTeramond:2024ikl, Binosi:2016nme} are set by 
different observables, such as hadron masses~\cite{Deur:2016cxb}.

\section{Determination of $\alpha_s(M^2_z)$ at EIC using the Bjorken sum rule \label{alpha_s@EIC}}
\vspace{-3mm}
We now turn to a possible determination of $\alpha_s(M^2_z)$ at EIC with the BJSR~\cite{Kutz:2024eaq}. 
First, we simulated doubly polarized electron-proton and electron-$^3$He DIS at the Comprehensive Chromodynamics Experiment (ECCE) detector~\cite{Adkins:2022jfp} of EIC, with 
double-tagging technique for neutron detection from $^3$He: both spectator protons from the $^3$He breakup are detected 
in the far-forward region. This selects nearly quasifree neutron scattering events, thereby suppressing nuclear uncertainties from 
$^3$He structure corrections.
We used three beam energy configurations for each two different ion beams, 
namely 5$\times$41, 10$\times$100, and 18$\times$275 GeV for e-p and 
the same for e-$^3$He except 18$\times$166 GeV for the highest energy. An integrated luminosity 10 fb$^{-1}$ for each 
configuration and 70\% polarization for both ion beams were assumed. Events were generated using DJANGOH~\cite{DJANGOH} 
and passed through a full GEANT4~\cite{GEANT4:2002zbu} simulation of ECCE to account for detector effects and electromagnetic radiative corrections. DIS cuts were then applied on 4-momentum transfer ($Q^2>2$~GeV$^2$), invariant mass ($W > \sqrt{10}$~GeV) and  inelasticity ($0.01<y<0.95$). 
Next, the inclusive double-polarization asymmetries were computed, from which $g_1(x, Q^2)$ was obtained.

\begin{wrapfigure}{R}{0.4\textwidth} 
  \vspace{-0.5cm}
   \includegraphics[width=0.4\textwidth]{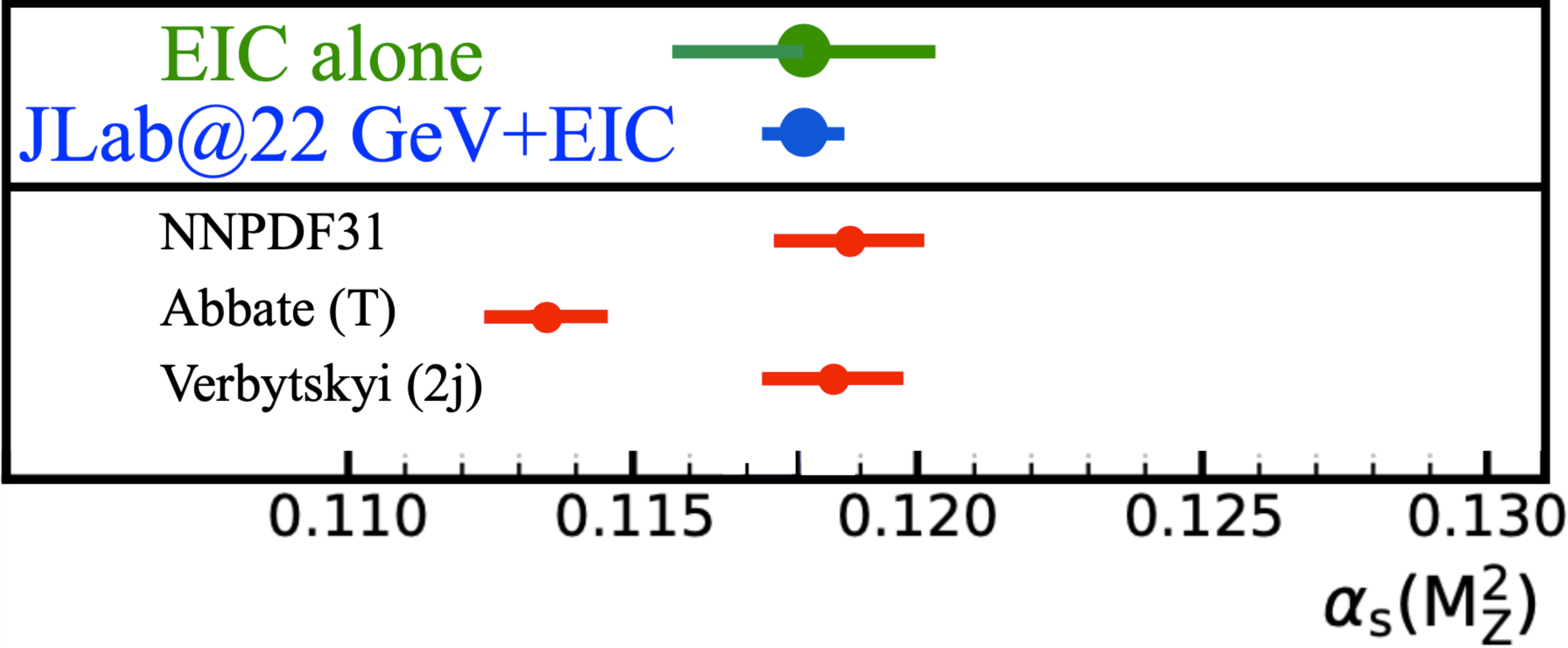}
    \vspace{-0.7cm}
  \caption{\small{Expected accuracy for $\alpha_s(M^2_z)$ from EIC and JLab@22 compared to the
  three most precise world data~\cite{ParticleDataGroup:2024cfk}.}} 
  \label{fig:alphas-22}
  \vspace{-0.5cm}
\end{wrapfigure}

To form $\Gamma_1^{p,n}$, $g_1^{p,n}$ was integrated over the $x$-range covered by the simulated data.
The unmeasured high-$x$ contribution was assessed using a parametrization of the $g_1$ world data~\cite{CLAS:2017qga}, 
while that at low-$x$ was obtained from the difference between the theoretical full $\Gamma_1^{p-n}$ and the simulated part.
The $Q^2$ evolution of $\Gamma_1^{p-n}$ is then fit using Eq.~(\ref{eq:BJSR}) in which $\alpha_s(Q^2)$ is itself expanded into
its $\beta$-series. This provides the QCD scale parameter $\Lambda_s$ and, from it, $\alpha_s(M_z)$. 
To assess the pQCD and $\beta$-series truncation uncertainties, we use N$^5$LO+twist-4
with $\alpha_s$ at N$^5$LO (i.e., $\beta_4$) \cite{Kniehl:2006bg} and take $|$N$^4$LO$-$N$^5$LO$|$/2 as the 
truncation error, where we use the $\beta_3$ order in the N$^4$LO estimate. There is an optimal range for the fit: 
too low $Q^2$ coverage provides greater sensitivity to $\alpha_s$ but prohibitive pQCD truncation and HT uncertainties; 
at too high $Q^2$, the sensitivity to $\alpha_s$ and reduction in the statistic uncertainty decline and do not balance the increase in
systematic uncertainty.
In all, the optimal fit covers $2.4 \leq Q^2 \leq 75$ GeV$^2$ and yields a relative uncertainty 
$\Delta \alpha_s/ \alpha_s = \pm 1.3\% = \pm (0.83\% \oplus 0.64\%)$, see Fig.~\ref{fig:alphas-22}, where 0.83\% is the fit uncertainty 
and 0.64\% comes from the truncation of the pQCD series. 
This precision is competitive with the current best $\alpha_s$ extractions from DIS global fits, $\sim$1.7\%~\cite{ParticleDataGroup:2024cfk}. 
%
%
Nevertheless, the precision can still be significantly improved with a complementary measurement at JLab@22, which we will now discuss.

\section{Measurement of $\alpha_s(M^2_z)$ and of its running at JLab@22}
\vspace{-3mm}
The energy range  and high luminosity of JLab@22 makes it ideally suited to extract $\alpha_s$ from the BJSR: 
At 22 GeV, one optimally balances sensitivity to $\alpha_s$ thanks to the relatively low $Q^2$ range covered, 
while keeping the pQCD truncation uncertainty small and the missing low-$x$ part of  $ \Gamma_1^{p-n}$ 
acceptable. (At lower JLab energies, the unmeasured low-$x$ piece prohibits accurate measurements of $ \Gamma_1^{p-n}$ 
for $Q^2 \gtrsim$ a few GeV$^2$.)
Negligible statistical uncertainties on $\Gamma_1^{p-n}$ are expected thanks to the high luminosity of JLab@22,
whose polarized DVCS and TMD programs will produce more than sufficient statistics for $\Gamma_1$, 
an inclusive and integrated observable.
For example, at 6 GeV, the polarized DVCS program produced 
statistical errors on $ \Gamma_1^{p-n}$ below 0.1\% at all $Q^2$~\cite{Deur:2014vea}. We thus assumed 0.1\% precision at 22 GeV for each $Q^2$ points, whose bin sizes increase exponentially to compensate the cross-section decrease with $Q^2$.
The experimental systematic uncertainty is expected to be about 5\%, coming from 
polarimetry (beam and target, 3\%),
target dilution/purity (NH$_3$ and $^3$He, 3\%),
nuclear corrections to obtain the neutron information (2\%),
the unpolarized structure function $F_1$ needed to form $g_1$ from the measured $A_1$ asymmetry (2\%), 
and radiative corrections (1\%).
The low-$x$ uncertainty is estimated as follows. For JLab $Q^2$ points also covered by EIC, PDF fits will be available down to the lowest $x$ 
covered by EIC. Thus, we use 10\% uncertainty on the low-$x$ part not measured at JLab@22 but covered by EIC, and 100\% for that not 
covered by the EIC. 
The five lowest $Q^2$ points of JLab@22 do not overlap with EIC. There, we use uncertainties ranging from 20\% to 100\% uncertainty, 
depending on the point proximity to the EIC $Q^2$-coverage. 
To extract  $\alpha_s(M^2_z)$ the simulated data are fit with Eq.~(\ref{eq:BJSR}), as we described in Section~\ref{alpha_s@EIC}. 
In that case, the optimal fit is found to range over 1$<Q^2<$8~GeV$^2$ and yields $ \Delta \alpha_s/\alpha_s \simeq 6.1\times10^{-3} $,
{\it viz}, more accurate than the current world data combined, see Fig.~\ref{fig:alphas-22}. 

\begin{figure}
\center
\includegraphics[width=0.6\textwidth]{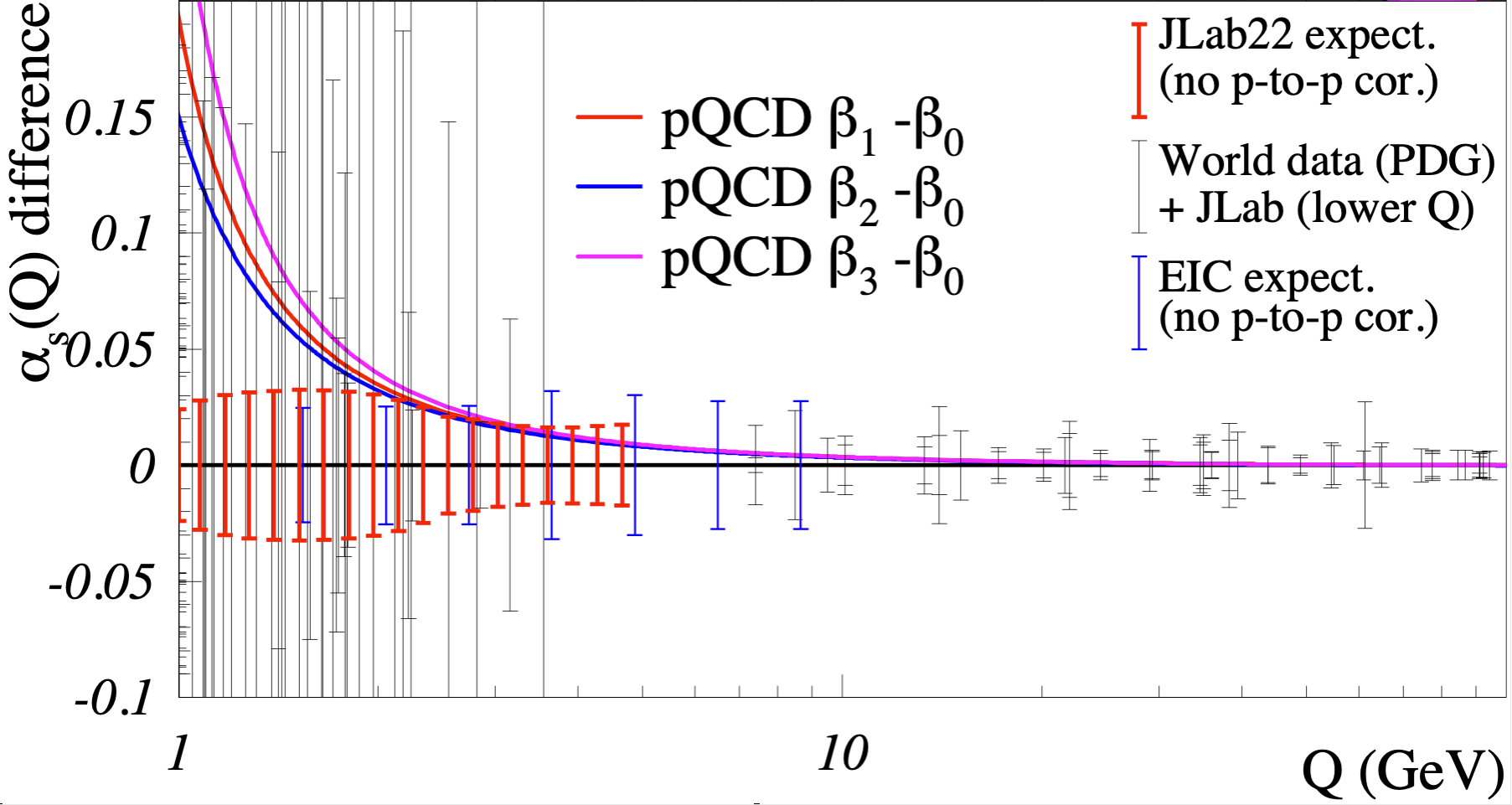}
\caption{\small{Contributions from higher loops on the running of $\alpha_s$: red ($\beta_1$, {\it i.e.}, NLO), blue ($\beta_2$, N$^2$LO) and magenta ($\beta_3$, N$^2$LO) lines. The black error bars centered on the horizontal 0-line show the uncertainties on $\alpha_s$ from the world data (Ref.~\cite{ParticleDataGroup:2024cfk}). The uncertainties expected from EIC and JLab@22GeV are shown by the blue and red error bars, respectively. JLab@22GeV+EIC can for the first time separate multi-loop effects, where non-QCD physics, including possible contributions of physics beyond the Standard Model, arise.}}
\label{fig:alpha_s_loops}
\end{figure}

The measurement also maps $\alpha_s$ over $1\lesssim Q \lesssim 9$~GeV, filling a region currently lacking point-to-point-accurate data, see Figs.~\ref{fig:alpha_s-all-scales} and \ref{fig:alpha_s_loops}. Comparing the point-to-point uncorrelated uncertainty of the
expected JLab@22 data with the effects of higher loops on the running of $\alpha_s(Q)$ reveals that the data can offer for the first time a 
 direct sensitivity to such effects.

\section{Conclusion}
\vspace{-3mm}
The Bjorken sum rule provides a relatively model-independent method to determine $\alpha_s$, since nonperturbative inputs are encapsulated in the precisely known axial charge $g_A$, the pQCD evolutions of moments are simpler than those of full structure functions, and the isovector combination 
suppresses sensitivity to gluon polarization. Furthermore, the extent of the unmeasurable low-$x$ contribution will soon be mitigated thanks to the EIC. Its high-precision doubly polarized DIS will allow the determination of $\alpha_s(M^2_z)$
with $\sim$1.3~\% relative accuracy. This is competitive with current DIS world‐data methods, and can be further reduced to 0.6~\% thanks to 
JLab@22. The BJSR is just one way to determine $\alpha_s$. Others, such as global PDF fits, will also be available at EIC and JLab and help
achieve the goal  of reaching the \textperthousand~ level in the upcoming decades~\cite{dEnterria:2022hzv}.

To determine the $Q^2$ behavior of $\alpha_s$, including in the nonperturbative domain, one can define $\alpha_s$ as an effective 
charge~\cite{Grunberg:1980ja}. Then, the Bjorken sum provides an especially well-suited observable that allows for an experimental
determination of $\alpha_s(Q)$  over four orders of magnitude in $Q$. The Bjorken sum rule (+pQCD) and the 
GDH sum rule complement the dataset for the higher and lower $Q$ regions, respectively. These data and their sum rule 
supplements are in excellent agreement with theoretical predictions of the effective charge from 
AdS/QCD~\cite{Brodsky:2010ur, deTeramond:2024ikl} and Schwinger-Dyson Equation/lattice QCD~\cite{Binosi:2016nme}, 
remarkably since these coupling calculations have no adjustable parameter.
JLab@22 will allow us to accurately map $\alpha_s(Q)$, covering $1\lesssim Q \lesssim 9$~GeV in particular, a region that currently
lacks data.
The JLab@22 data will be directly sensitive to the effects of higher loops on the running of $\alpha_s(Q)$, thereby testing pQCD in a novel way and providing a new window on possible  physics beyond the Standard Model.

{\bf Acknowledgements}
This material is based upon work supported by the
U.S.\ Department of Energy, Office of Science, Office of Nuclear Physics, contract DE-AC05-06OR23177.
The author thanks S.~J.~Brodsky and G.~F.~de T\'eramond for enlightening discussion on $\alpha_s$ and the  organizers of the ``QCD at the Extremes'' workshop, H. Jung, K.Kutak, N. Raicevic and S. Taheri-Monfared for the invitation to present this research.

\end{document}